\documentclass[a4paper]{article}
\usepackage{amsmath}
\usepackage{amsbsy,bm}
\usepackage{CJK}
\usepackage{graphicx}
\usepackage{color}
\usepackage{indentfirst}
\usepackage{hyperref}
\begin{document}
%
%
\title{Central compact objects, superslow X-ray pulsars, gamma-ray bursts:
 do they have anything to do with magnetars?}
\author{H. Tong\thanks{Xinjiang Astronomical Observatory, Chinese Academy of Sciences, Urumqi, Xinjiang 830011,China; 
       tonghao@xao.ac.cn}, 
       W. Wang\thanks{National Astronomical Observatories, Chinese Academy of Sciences, Beijing 100012, China;
       wangwei@bao.ac.cn}}
\date{2014.6}
\maketitle
%
%
%
\begin{abstract}
Magnetars and many of the magnetar-related objects are summarized together and discussed. It is shown that
  there is an abuse of language in the use of ``magnetar''. Anomalous X-ray pulsars and
  soft gamma-ray repeaters are well-known magnetar candidates. The current so called anti-magnetar
  (for central compact objects), accreting magnetar (for superslow X-ray pulsars in high mass X-ray
  binaries), and millisecond magnetar (for the central engine of some gamma-ray bursts),
  they may not be real magnetars in present understandings.
  Their observational behaviors are not caused by the magnetic energy.
  Many of them are just neutron stars with strong surface dipole field.
  A neutron star plus strong dipole field is not a magnetar.
  The characteristic parameters of the neutron stars for the central engine of some gamma-ray bursts
  are atypical from the neutron stars in the Galaxy.
  Possible signature of magnetic activities
  in accreting systems are discussed, including repeated bursts and a hard X-ray tail.
  China's future hard X-ray modulation telescope may contribute to finding some
  magnetic activities in accreting neutron star systems.
\end{abstract}
%
%
\section{Introduction}
Magnetar is a new kind of pulsars compared with rotation-powered and accretion-powered pulsars (Duncan \& Thompson 1992).
The star's radiations (both persistent and burst) are from their magnetic energy (Thompson \& Duncan 1995, 1996).
Observationally, anomalous X-ray pulsars (AXPs) and soft gamma-ray repeaters (SGRs) are assumed to be magnetars (Mereghetti 2008).
The release of magnetic energy can be in all wavelength, including radio, optical/infrared,
soft and hard X-ray etc (Mereghetti 2008; Olausen \& Kaspi 2014).
Besides AXPs and SGRs, there exist other magnetar-related objects. X-ray dim isolated neutron stars are nearby sources
with high surface dipole field ($\sim 10^{13} \,\rm G$),
simple blackbody spectrum and without radiative activities (Turolla 2009; Tong et al. 2010, 2011).
They are supposed to be dead magnetars (Vigano et al. 2013).
Central compact objects are compact stars lying at the center of the supernova remnant (De Luca 2008).
Their small hot spot in X-rays may require the crustal field up to magnetar strength (Shabaltas \& Lai 2012).
While timing observations only find a very small surface dipole field
($\sim 10^{10} \,\rm G$, Halpern \& Gotthelf 2010).
Therefore, they are dubbed anti-magnetars (Gotthelf \& Halpern 2008).

Superslow X-ray pulsars are special sources in high mass X-ray binaries (Wang 2013a).
Their spin period is longer than 1000 seconds.
Large period derivatives are also observed (both negative and positive).
Their very long spin period and large period derivative may require a magnetar strength surface dipole field. Therefore,
they are supposed to be accreting magnetars.
The central engine of some gamma-ray burst may be a millisecond magnetar (Usov 1992; L$\rm{\ddot{u}}$ \& Zhang 2014).
If the surface dipole field is higher than $10^{15} \,\rm G$, the rotational energy
may be released in a very short timescale, thus result in a gamma-ray burst.

However, for these so called magnetar-related objects (anti-magnetar, accreting magnetar, millisecond magnetar),
their radiations are not powered by the magnetic energy. According to present observations, they are just neutron stars
with strong surface dipole field. Current understandings of magnetars tell us that
a neutron star plus strong surface dipole field is not a magnetar.
The names of these objects are improper and misleading.
In the following, we will first show what is a magnetar. Then the essence of various magnetar-related objects
will be analyzed.

\section{Magnetars and related objects}

\subsection{Magnetar observational behaviors}

\begin{figure}[!ht]
\centering
 \includegraphics[width=0.7\textwidth]{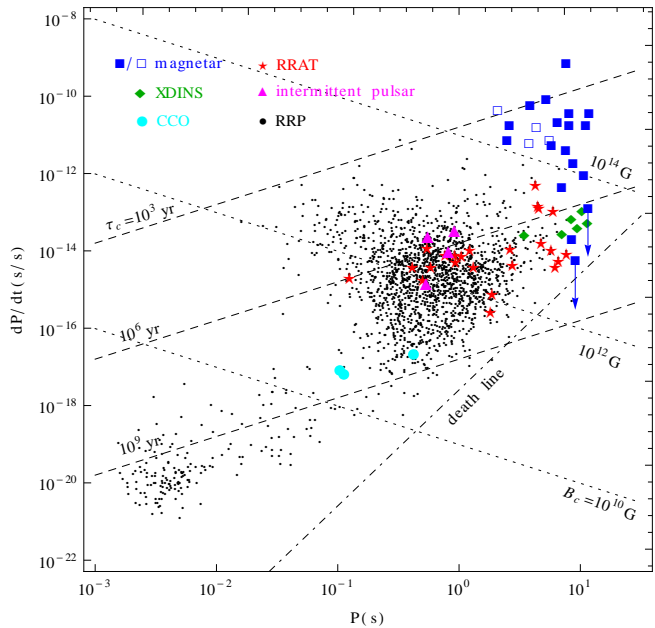}
\caption{Pulsars and magnetars on the period-period derivative diagram.
Blue squares are magnetars (empty squares are radio-loud magnetars)
(from McGill magnetar catalog: http://www.physics.mcgill.ca/$\sim$pulsar/magnetar/main.html).
Green diamonds are X-ray dim isolated neutron stars
(from Kaplan \& van Kerkwijk 2011 and references therein),
cyan circles are central compact objects (from Halpern \& Gotthelf 2010; Gotthelf et al. 2013),
red stars are rotating radio transients
(from http://astro.phys.wvu.edu/rratalog/),
magenta triangles are intermittent pulsars
(From Lorimer et al. 2012 and references therein; Surnis et al. 2013),
and black dots are rotation powered pulsars (including normal pulsars and millisecond pulsars,
which dominate the pulsar population)
(from ATNF: http://www.atnf.csiro.au/research/pulsar/psrcat/).}
\label{fig_PPdot}
\end{figure}

AXPs and SGRs are the most likely magnetar candidates.
Currently, there are more than 20 AXPs and SGRs discovered\footnote{See the McGill magnetar catalog
http://www.physics.mcgill.ca/$\sim$pulsar/magnetar/main.html}.
Their multiwavelength observations have been reviewed by many authors (Woods \& Thompson 2006; Kaspi 2007;
Mereghetti 2008; Rea \& Esposito 2011; Olausen \& Kaspi 2014). The salient features can be summarized
as follows.
\begin{enumerate}

  \item Persistent and burst emissions. Many of the anomalous X-ray pulsars and soft gamma-ray repeaters
  have persistent X-ray luminosity higher than their rotational energy loss rate (e.g.,
  Figure 13 in Olausen \& Kaspi 2014; Figure 1 in Tong et al. 2013). At the same time, they show various
  kinds of burst, outburst and flares (Woods \& Thompson 2006; Kaspi 2007; Mereghetti 2008; Rea \& Esposito 2011).
  These two aspects are the key reasons to employ the magnetar model, which provides a new energy reservoir
  (Thompson \& Duncan 1995, 1996).

  \item Timing behaviors. From their distribution on the $P - \dot P$ diagram of pulsars
 ( see Figure \ref{fig_PPdot}),
  magnetars have pulsation period clustered around $2-12$ seconds.
  At the same time, their period derivatives span a broader range (from $10^{-14}$ to $10^{-10}$).
  This scatter of period derivative results in a scatter of the magnetars' characteristic magnetic field
  (from less than $10^{13} \,\rm G$ to higher than $10^{15} \,\rm G$), and characteristic
  age (from $10^3 \,\rm yr$ to more than $10^{7} \,\rm yr$).
  Therefore, a long pulsation period is a common characteristic of magnetars (not the period derivative or dipole field).

  \item Multiwavelength spectral properties and variabilities.
  Both the spindown rate and X-ray emissions of magnetars change with time
  (e.g., SGR J1745$-$2900 (Kaspi et al. 2014; Tong 2014); see Tong \& Xu 2014 for more information).
  Variations in radiation and timing are also observed in radio observations of magnetars
  (Camilo et al. 2007; Levin et al. 2012). In the magnetar model, both the radiation and braking torque are
  dominated by magnetic field (Thompson et al. 2002; Tong et al. 2013). Therefore, it is natural that there
  is a correlation between radiative and timing events of magnetars (Tong \& Xu 2014). Compared with the X-ray radiation
  of X-ray dim isolated neutron stars and central compact objects, magnetars have a significant nonthermal radiation.
  There is a power law component in the soft X-ray band (power law index $\sim 3$, Rea et al. 2008), another
  power low component in the hard X-ray range (power law index $\sim 1$, Gotz et al. 2006; Wang et al. 2014),
  a flat radio spectra (Camilo et al. 2006), and nonthermal optical emissions (Wang et al. 2006).
\end{enumerate}

These features make magnetars different from high magnetic field pulsars, X-ray dim isolated
neutron stars and rotating radio transients (although they overlap on the period-period derivative diagram,
as can be seen from Figure \ref{fig_PPdot}).
They are all caused by the magnetic activities of the central neutron star.
The putative neutron star can have so many magnetic activities because it has a very strong multipole field
(either crustal or magnetospheric,  Thompson et al. 2002; Beloborodov 2009; Tong et al. 2013; Vigano et al. 2013).

\subsection{Low magnetic field magnetars and anti-magnetars}

The discovery of low magnetic field magnetar SGR 0418+5729 challenged the traditional magnetar model directly (Rea et al. 2010).
Its characteristic magnetic field is less than $7.5\times 10^{12} \,\rm G$ and characteristic age larger than
$2.4\times 10^7 \,\rm yr$ (Rea et al. 2010). It may be an aged magnetar (Turolla et al. 2011) or
a normal magnetar with very small inclination angle (Tong \& Xu 2012).
The existence of low magnetic field magnetars demonstrate
that the presence of strong multipole field is responsible for the magnetar activities (anomalous X-ray luminosity
and/or bursts).

The X-ray luminosity of central compact objects is similar to that of the quiescent state of transient magnetars
(Halpern \& Gotthelf 2010). Their small hot spot (with radius $\sim 1\,\rm km$ and temperature $\sim 0.5 \,\rm keV$)
may require magnetar-strength crustal field (due to anisotropic conduction, Shabaltas \& Lai 2012).
However, timing observations found that they have very small period derivative ($\sim 10^{-17}$, Halpern \& Gotthelf 2010;
Gotthelf et al. 2013). This may indicate that they are born as slow rotators and with very weak surface dipole field
($\sim 10^{10}\,\rm G$, Halpern \& Gotthelf 2010). Therefore they are dubbed as anti-magnetars (Gotthelf \& Halpern 2008).
The problem in the anti-magnetar prescription is that a weak surface dipole field
in combination with a magnetar-strength crustal field is the same as low magnetic field magnetars (Turolla et al. 2011).
Furthermore, central compact objects (which are still associated with
supernova remnants) may be much younger than low magnetic field magnetars.
Then, in the anti-magnetar model, central compact objects should be more active
than low magnetic field magnetars. However, central compact objects show very weak variabilities (if any) and
no significant nonthermal X-ray radiations (Halpern \& Gotthelf 2010). In conclusion, at present there is no
strong evidence for the existence of strong crustal field in central compact objects\footnote{There are many literatures
discussing the origin/evolution of central compact objects' weak surface dipole field. These literatures
can not contribute to the solution of the small hot spot problem (this is the radiative properties of central compact objects).}.

Central compact objects may be an independent manifestation of young neutron stars, in addition to rotation-powered pulsars
(e.g., Crab and Vela pulsars) and magnetars. The name ``anti-magnetar'' may be improper and misleading. An independent name
may be better, e.g., they may be called ``magninos'' (which was also proposed
by Gotthelf\footnote{http://xmm.esac.esa.int/external/xmm\_science/workshops/2011symposium/talks/Gotthelf\_TopicD.pdf}).

\subsection{Accreting magnetar model for superslow X-ray pulsars}

Superslow X-ray pulsars are a special class of high mass X-ray binaries. They have rotation
period longer than 1000 seconds (Wang 2013a, also see Fig. 2). Both their long spin period and large period derivatives
may require that their surface dipole fields are as large as $10^{14}- 10^{15} \,\rm G$ (by employing
similar accretion torque to the disk accretion case (Lai 2014), see Wang 2013b and references therein).
At the same time, possible cyclotron lines are also observed in these sources (e.g., 4U 2206+54, Wang 2009;
2S 0114+61, Bonning et al. 2005, Wang et al. 2011). The presence of harmonics implies that these
cyclotron lines are more likely due to electron origin (Potekhin 2010). However, the cyclotron features
may be formed above the neutron star surface (e.g., at five stellar radius, and at this radius
the remaining magnetic field is just the dipole field). Then the corresponding
surface dipole field will also be larger than $10^{14}\,\rm G$. Therefore, superslow X-ray pulsars
are often named as accreting magnetars (Wang 2013a, 2013b).

However, all the required magnetic field is just the surface dipole field.
And the X-ray emissions are dominated by accretion power (e.g., Wang 2009).
Therefore based on present observations,
they are just accreting high magnetic field neutron stars\footnote{Considering the differences between
wind accretion torque and disk accretion torque, the surface dipole field of superslow X-ray pulsars
may just be about $10^{12} \,\rm G$ (Shakura et al. 2012).}.

Similar things also happen in high magnetic field pulsars (McLaughlin et al. 2003; Ng \& Kaspi 2010).
These sources may have higher surface dipole field ($\sim 10^{14} \,\rm G$, from timing observations).
At the same time, their radio and X-ray emissions are more likely powered by the rotational energy.
Therefore, they are not magnetars and they are named as high magnetic field pulsars.

Some magnetar activities are observed in high magnetic field pulsars (Gavriil et al. 2008).
Considering superslow X-ray pulsars may have high surface dipole field, finding
some magnetar activities in these sources is also promising.
Soft gamma-ray repeater like burst is different from the type I and type II burst
in X-ray binaries (Gavriil et al. 2002). Therefore, the most convincing evidence
of mangetar activity is finding some repeated bursts from some superslow X-ray pulsars
(e.g., one SGR-like burst by Torres et al. 2012). The release of magnetic energy may result in a rather flat
hard X-ray spectrum (power index $\sim 1$) and a cutoff at about $100\,\rm keV$ (Wang et al. 2014).
Around $100\,\rm keV$, the emission due to accretion may be greatly suppressed (Wang et al. 2014).
Then, an additional hard X-ray tail in superslow X-ray pulsars may be due to magnetic energy release.

In one accreting magnetar candidate 2S 0114+65, a possible hard X-ray tail was observed (Wang 2011).
In addition, the hard X-ray tail are only detected when the column density
is very low based on the combined soft and hard X-ray spectral studies on 2S 0114+65.
For optimal conditions, a magnetar corona may be formed around the central
neutron star (Beloborodov \& Thompson 2007). The hard X-ray tail in 2S 0114+65 may be
generated by Compton process in the magnetar corona (the soft X-ray photons as the seed photons).
If the column density is from the accretion origin, a high column density may suppress the
the formation of such a corona. A second possibility is that the soft X-ray emission will
be absorbed significantly when the column density is high. There will be fewer soft X-ray
photons to be scattered in the magnetar corona. Therefore, the hard X-ray tail can not be
formed when the column density is high.
More observations are needed in order to confirm the existence of such a hard
X-ray tail and its magnetic origin. The hard X-ray modulation telescope (HXMT, Wang et al. 2014)
will have better performance at $\sim 100\,\rm keV$. Combining with other telescopes,
finding some magnetar activities in superslow X-ray pulsars (and other high mass X-ray binaries)
may be possible. Only at that time, one can say that some magnetar activities are observed in
accreting neutron stars. At present, the radiation and timing properties of superslow X-ray pulsars
are just the characteristics of accreting high magnetic field neutron stars.
A summary of isolated magnetars and superslow X-ray pulsars is given in Table \ref{tab_magnetar}.

\begin{figure}[!t]
\centering
\includegraphics[angle=0,width=0.8\textwidth]{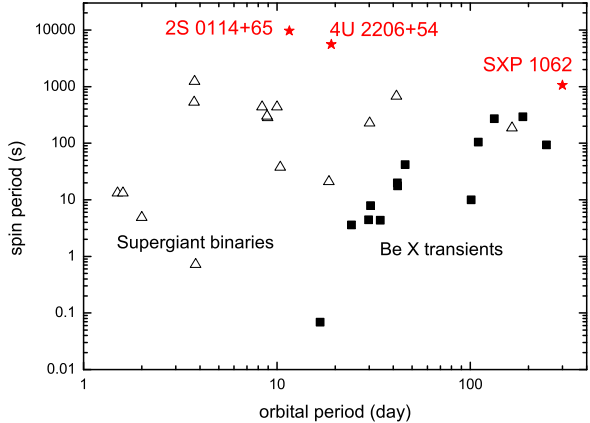}
\caption{The spin period -- orbital period diagram for the accreting X-ray pulsars in high mass X-ray binaries.
The red stars show the three accreting magnetar candidates.
The data point for 2S 0114+65 is taken from Wang (2011); the data point for 4U 2206+54 is taken from Wang (2009; 2013b);
the data point of SXP 1062 is taken from Haberl et al. (2012).
The data points of supergiant binaries (triangles) and Be X-ray transients (squares)
are taken from Bildsten et al. (1997) and Sidoli (2011).}
\end{figure}

\begin{table}[!t]
\caption{Present understanding of isolated magnetars and superslow X-ray pulsars}
\scriptsize
\centering
\begin{tabular}{lcl}
\hline \hline
Observational manifestation & AXPs/SGRs & superslow X-ray pulsars \\\hline

Spin period (s) &  2--12 & $>$ 1000 \\\hline

Spin period derivative (s/s) & $10^{-14}-10^{-10}$  & $10^{-7}-10^{-6}$ \\
& & (both positive and negative)\\\hline

Characteristic age (yr) & $10^{3}-10^7$ & $10^4-10^6$ \\\hline

X-ray spectrum & 0.1-10 keV: $kT\sim 0.5 \,\rm keV$, $\Gamma \sim 3$  & Power-law  $\Gamma \sim 1.5-2$\\
 & $>10$ keV: $\Gamma\sim 1$  &   (with high energy cutoff) \\\hline

X-ray luminosity (erg/s)   & Quiescence $10^{33}-10^{35}$  & $10^{34}-10^{37}$ \\
 (1-100 keV)  & Bursts $>10^{37}$  &  \\\hline

Energy power  & magnetic field  & wind-fed accretion \\\hline

Theoretical model & isolated magnetar & accreting high-B neutron star \\

& & (may have magnetar activities) \\
\hline

\end{tabular}\label{tab_magnetar}
\end{table}

\subsection{Millisecond magnetars as the central engine of some gamma-ray bursts}

When magnetar was first proposed, explaining gamma-ray bursts was one of the aims
(Duncan \& Thompson 1992; Usov 1992). For a neutron star with polar surface dipole field
as high as $3\times 10^{15} \,\rm G$ and angular velocity $10^{4} \,\rm s^{-1}$, the star's
total rotational energy is $5\times 10^{52} \,\rm erg$ (Usov 1992). Most of these energies
will be released during a very short time. Assuming magnetic dipole spindown, the typical
spin-down time scale will be about $20 \,\rm s$ (Usov 1992). The spindown luminosity will
be about $2\times 10^{51} \,\rm erg \, s^{-1}$ (Usov 1992). Therefore, a rapid rotating
high magnetic field neutron star can be the central engine of some gamma-ray bursts (Usov 1992;
L$\rm{\ddot{u}}$ \& Zhang 2014). This scenario is the currently so called ``millisecond magnetar''
model for the central engine of some gamma-ray bursts (L$\rm{\ddot{u}}$ \& Zhang 2014 and references
therein).

The problem with the ``millisecond magnetar'' name is that: the required energy is the rotational
energy (it is not the magnetic field energy\footnote{Magnetic energy may also power some late activities
of gamma-ray bursts, e.g., X-ray flares (Dai et al. 2006).}); the required magnetic field
is the surface dipole field (it is not the multipole field or toroidal field as the case of magnetars).
Therefore, the requirement is just a rapid rotating neutron star with strong surface dipole field.
Furthermore, the required surface dipole field increases with the required initial rotation period
(e.g., Figure 6 in L$\rm{\ddot{u}}$ \& Zhang 2014). The extreme cases are (for different sources,
Table 2 in L$\rm{\ddot{u}}$ \& Zhang 2014):
for an initial rotation period of $1.39 \,\rm ms$, the required surface dipole field is
$3.6\times 10^{14} \,\rm G$; while for an initial period of $59\,\rm ms$, the required surface dipole
field is $10^{17} \,\rm G$. These parameters are atypical from our current knowledge of magnetars
and related pulsar-like objects. Magnetars may be born as fast rotators (e.g., with rotation
period of several milliseconds, Duncan \& Thompson 1992).
This is why they can have very high surface dipole field (and higher multipole field, Duncan \& Thompson 1992).
While the existence of central compact objects require the existence of neutron stars born as
slow rotators and weakly magnetized at birth (e.g., with rotation period of $100\,\rm ms$
and surface dipole field $10^{10}\,\rm G$, Halpern \& Gotthelf 2010; Gotthelf et al. 2013).
Therefore, the characteristic parameters of the neutron stars for the central engine of gamma-ray bursts
are atypical from the neutron stars in the Galaxy.

\section{Discussions and conclusions}

Based on the study of 20 AXPs and SGRs, the key feature of magnetars is
their variability compared with normal rotation-powered pulsars. Among the variabilities of magnetars,
the anomalous X-ray luminosity and recurrent soft gamma-ray burst are two representives.
The underlying physics of magnetar's activities is that they are neutron stars with strong multipole
field (Thompson et al. 2002; Beloborodov 2009; Tong et al. 2013; Vagino et al. 2013).
A strong surface dipole field is not necessary (Tong et al. 2013).
This point is demonstrated clearly by the existence of low magnetic field magnetars (Rea et al. 2010).
The anti-magnetar model for central compact objects is similar to the case of low magnetic field magnetars
(Halpern \& Gotthelf 2010; Turolla et al. 2011). The lack of variability in central compact objects
argues against the anti-magnetar model.
It may be an independent manifestation of isolated neutron stars. Therefore, the anit-magnetar name is improper
and it should have an independent name.

An isolated neutron star with high surface dipole field (e.g., $10^{14} \,\rm G$)
may not be a magnetar.
It could be just a high magnetic field pulsar (Ng \& Kaspi 2010), whose radio and X-ray radiations
are powered by the rotational energy. An accreting high magnetic field neutron star may explain
the radiative and timing of superslow X-ray pulsars in some high mass X-ray binaries (Wang 2013a, b).
The X-ray emission and timing variabilities of these superslow X-ray pulsars
are dominated by accretion power. So they may not be defined as a magnetar family -- accreting magnetar based on present observations.
In addition, The birth of a rapidly rotating and strongly magnetized
neutron star may result in a comic gamma-ray burst (Usov 1992). The burst is powered by
the rotational energy of the neutron star. Therefore, this special neutron star may not be named as a millisecond magnetar,
but just a millisecond high magnetic field neutron star.  Meanwhile, the required parameters in the model of gamma-ray bursts
are atypical from the neutron stars in the Galaxy.

Magnetars are just one kind of neutron stars. Therefore, they should also be found in
binary systems (Woods \& Thompson 2006). The key problem is whether the magnetic actives can be
identified in the presence of accretion power. From the experience of isolated magnetars,
magnetic energy may result in a flat hard X-ray spectrum. An additional hard X-ray
component in the high energy end may from the magnetic origin. The smoking gun evidence
is finding some repeated bursts from accreting systems (as have been done in the case of
high magnetic field pulsars, Gavriil et al. 2008). China's future hard X-ray modulation
telescope may contribute to these aspects.

In summay, according to the current understanding, anomalous X-ray pulsars and soft gamma-ray repeaters
are likely to be magnetars (e.g., magnetic activities dominate the star's observational behaviors).
Some magnetic activities are discovered in one high magnetic field pulsar. No significant magnetic
activities are found in central compact objects. Finding some magnetic activities in superslow
X-ray pulsars is promising. The currently so called anti-magnetar (for central compact objects),
accreting magnetar (for superslow X-ray pulsars), and millisecond magnetar (for the central engine
of some gamma-ray bursts) are improper and misleading. There is an abuse of language in the use of
``magnetar''. A neutron star plus strong surface dipole field is not a magnetar.

\section*{Acknowledgments}
The authors would like to thank R.X.Xu and J.L.Han for discussions.
H.Tong is supported by NSFC (11103021), West Light Foundation of CAS (LHXZ201201), Xinjiang Bairen project,
and Qing Cu Hui of CAS.

\end{document}